\documentclass{article}

\usepackage[utf8]{inputenc}

\usepackage[english]{babel}
\usepackage{subeqnarray}
\usepackage{url}

\usepackage{graphicx,psfrag}
\usepackage{fixltx2e,amsmath}
\usepackage{xcolor}
\usepackage{amssymb}
\usepackage[final]{pdfpages}
\usepackage{bbold}
\usepackage{dsfont}
\usepackage{pgfplots}
\usepackage{subfig}
\usepackage{algorithm}
\usepackage{algorithmic,placeins}
\usepackage[font=small,labelfont=bf]{caption}

\usepackage[font={footnotesize}]{caption}

\usepackage[utf8]{inputenc}
\usepackage[T1]{fontenc}
\usepackage{xcolor,amsmath,amssymb}
\usepackage{geometry}
\usepackage{graphicx}
\usepackage[sort&compress,numbers]{natbib}
\usepackage{enumerate}
\usepackage{wasysym}
\usepackage[textsize=footnotesize]{todonotes}

\usepackage{hyperref}
\hypersetup{
    colorlinks=true,
    linkcolor=red,
    filecolor=magenta,      
    urlcolor=cyan,
}

\usepackage{pgfplots}

\definecolor{darkblue}{rgb}{0,0,0.6}
\definecolor{darkred}{rgb}{0.6,0,0}

\usepackage{hyperref}

\newcommand{\beq}{\begin{equation}} \newcommand{\eeq}{\end{equation}}

\newcommand\be{\begin{equation}}
\newcommand\bea{\begin{eqnarray} \nonumber }
\newcommand\ee{\end{equation}}
\newcommand\eea{\end{eqnarray}}

\usepackage{authblk}

\usepackage[bitstream-charter]{mathdesign}

\usetikzlibrary{pgfplots.groupplots}

\graphicspath{{figures/}} 

\begin{document}
\title{Radical Complexity}
\author{Jean-Philippe Bouchaud, CFM \& Acad\'emie des Sciences}
\maketitle
\begin{abstract}
This is an informal and sketchy review of six topical, somewhat unrelated subjects in quantitative finance: rough volatility models; random covariance matrix theory; copulas; crowded trades; high-frequency trading \& market stability; and ``radical complexity'' \& scenario based (macro)economics. Some open questions and research directions are briefly discussed.
\end{abstract}

\section{From Random Walks to Rough Volatility}

Since we will never really know {\it why} the prices of financial assets move, we should at least make a model of {\it how} they move. This was the motivation of Bachelier in 1900, when he wrote in the introduction of his thesis that {\it contradictory opinions in regard to [price] fluctuations are so diverse that at the same instant buyers believe the market is rising and sellers that it is falling}. He went on to propose the first mathematical model of prices: the Brownian motion. He then built an option pricing theory that he compared to empirical data available to him --- which already revealed, quite remarkably, what is now called the volatility smile! 

After 120 years of improvements and refinements, we are closing in on a remarkably realistic model, which reproduces almost all known stylized facts of financial price series. But are we there yet? As Beno\^\i t Mandelbrot once remarked: {\it In economics, there can never be a ``theory of everything''. But I believe each
attempt comes closer to a proper understanding of how markets behave.} In order to close the gap, and justify the modern mathematical apparatus that has slowly matured, we will need to understand the interactions between the behaviour of zillions of traders --- each with his or her own investment style, trading frequency, risk limits, etc. and the price process itself. Interestingly, recent research strongly suggests that markets self organise in subtle way, as to be poised at the border between stability and instability. This could be the missing link --- or the holy grail -- that researchers have been looking for. 

For many years, the only modification to Bachelier's proposal was to consider that log-prices, not prices themselves, are described by a Brownian motion. Apart from the fact that this modification prevents prices from becoming negative, none of the flaws of the Bachelier model were seriously tackled. Notwithstanding, the heyday of Brownian finance came when Black \& Scholes published their famous 1973 paper, with the striking result that perfect delta-hedging is possible. But this is because, in the Black-Scholes world, price jumps are absent and crashes impossible. This is of course a very problematic assumption, specially because the fat-tailed distribution of returns had been highlighted as soon as 1963 by Mandelbrot --- who noted, in the same paper, that {\it large changes tend to be followed by large changes, of either sign, and small changes tend to be followed by small changes}, an effect now commonly referred to as ``volatility clustering'', and captured by the extended family of GARCH models.      

It took the violent crash of October 1987, exacerbated by the massive impact of Black-Scholes' delta-hedging, for new models to emerge. The Heston model, published in 1993, is among the most famous post-Black-Scholes models, encapsulating volatility clustering within a continuous time, Brownian motion formalism. However, like GARCH, the Heston model predicts that volatility fluctuations decay over a single time scale --- in other words that periods of high or low volatility have a rather well defined duration. This is not compatible with market data: volatility fluctuations have no clear characteristic time scale; volatility bursts can last anything between a few hours and a few years. 

Mandelbrot had been mulling about this for a long while, and actually proposed in 1974 a model to describe a very similar phenomenon in turbulent flows, called ``multifractality''. He adapted his theory in 1997 to describe currency exchange rates, before Emmanuel Bacry, Jean-Francois Muzy \& Jean Delour formulated in 2000 a more convincing version of the model, which they called the {\it Multifractal Random Walk} (MRW) \cite{MRW}. With a single extra parameter (interpreted as a kind of volatility of volatility), the MRW captures satisfactorily many important empirical observations: fat-tailed distribution of returns, long-memory of volatility fluctuations. In 2014, Jim Gatheral, Thibault Jaisson and Mathieu Rosenbaum introduced their now famous ``Rough Volatility'' model \cite{Rough}, which can be seen as an extension of the MRW with an extra parameter allowing one to tune at will the roughness of volatility, while it is fixed in stone in the MRW model. And indeed, empirical data  suggests that volatility is slightly less rough than what the MRW posits. Technically, the Holder regularity of the volatility is $H=0$ in the MRW and found to be $H \approx 0.1$ when calibrated within the Rough Volatility specification.  

The next episode of the long saga came in 2009 when Gilles Zumbach noticed a subtle, yet crucial aspect of empirical financial time series: they are not statistically invariant upon time reversal \cite{Zumbach}. Past and future are not equivalent, whereas almost all models to that date, including the MRW, did not distinguish past from future. More precisely, past price trends, whether up or down, lead to higher future volatility, but not the other way round. In 2019, following some work by Pierre Blanc, Jonathan Donier and myself \cite{QHawkes}, Aditi Dandapani, Paul Jusselin, Mathieu Rosenbaum proposed to describe financial time series with what they called a ``Quadratic Rough Heston Model'' \cite{QRough}, which is a synthesis of all the ideas reviewed above. It is probably the most realistic model of financial price series to date. In particular, it provides a natural solution to a long standing puzzle, namely the joint calibration of the volatility smile of the S\&P 500 and VIX options, which had eluded quants for many years \cite{Guyon}. The missing ingredient was indeed the Zumbach effect \cite{VIX}.  

Is this the end of the saga? From a purely engineering point of view, the latest version of the Rough Volatility model is probably hard to beat. But the remaining challenge is to justify how this particular model emerges from the underlying flow of buy and sell trades that interacts with market makers and high frequency traders. Parts of the story are already clear; in particular, as argued by Jaisson, Jusselin \& Rosenbaum in a remarkable series of papers, the Rough Volatility model is intimately related to the proximity of an instability \cite{Rosenbaum1} (see also \cite{Hardiman}), that justifies the rough, multi-timescale nature of volatility. But what is the self-organising mechanism through which all markets appear to settle close to such a critical point? Could this scenario allow one to understand why financial time series all look so much alike --- stocks, futures, commodities, exchange rates, etc. share very similar statistical features, in particular in the tails. Beyond being the denouement of a 120 years odyssey, we would be allowed to believe that the final model is not only a figment of our mathematical imagination, but a robust, trustworthy framework for risk management and derivative pricing.

\section{Random Matrix Theory to the Rescue}

Harry Markowitz famously quipped that diversification is the only free lunch in finance. This is nevertheless only true if correlations are known and stable over time. Markowitz’ optimal portfolio offers the best risk-reward tradeoff, for a given set of predictors, but requires the covariance matrix -- of a potentially large pool of assets -- to be known and representative of the future realized correlations. 
The empirical determination of large covariance matrices is however fraught with difficulties and biases. Interestingly, the vibrant field of ``Random Matrix Theory” has provided original solutions to this big data problem, and suggests droves of possible applications in econometrics, machine learning or other large dimensional models.  

But even for the simplest two-asset bond/equity allocation problem, the knowledge of the forward looking correlation has momentous consequences for most asset allocators in the planet. Will this correlation remain negative in the years to come, as it has been since late 1997, or will it jump back to positive territories? But compared to volatility, our understanding of correlation dynamics is remarkably poor and, surprisingly, the hedging instruments allowing one to mitigate the risk of bond/equity correlation swings are nowhere as liquid as the VIX itself. 

So there are two distinct problems in estimating correlation matrices. One is lack of data, the other one is time non-stationarity. Consider a pool of $N$ assets, with $N$ large. We have at our disposal $T$ observations (say daily returns) for each of the $N$ time series. The paradoxical situation is this: even though each individual off-diagonal covariance is accurately determined when $T$ is large, the covariance matrix as a whole is strongly biased unless $T$ is much larger than $N$ itself. For large portfolios, with $N$ a few thousands, the number of days in the sample should be in the tens of thousands – say 50 years of data. This is simply absurd: Amazon and Tesla did not even exist 25 years ago. Maybe use 5 minutes returns then, increasing the number of data points by a factor 100? Yes, except that 5 minute correlations are not necessarily representative of the risk of much lower frequency strategies, with other possible biases creeping in the resulting portfolios.   

So in what sense are covariance matrices biased when $T$ is not very large compared to $N$? The best way to describe such biases is in terms of eigenvalues. One finds that the smallest eigenvalues are way too small and the largest eigenvalues are too large. This results, in the Markowitz optimization program, in a substantial over-allocation on combination of assets that happened to have a small volatility in the past, with no guarantee that this will persist looking forward. The Markowitz construction can therefore lead to a considerable under-estimation of the realized risk in the next period. 

Out-of-sample results are of course always worse than expected, but Random Matrix Theory (RMT) offers a guide to (partially) correct these biases when $N$ is large. In fact, RMT gives an optimal, mathematically rigorous, recipe to tweak the value of the eigenvalues so that the resulting “cleaned” covariance matrix is as close as possible to the “true” (but unknown) one, in the absence of any prior information \cite{Bun}. Such a result, first derived by Ledoit and Péché in 2011 \cite{LP}, is already a classic and has been extended in many directions. The underlying mathematics, initially based on abstract ``free probabilities”, are now in a ready-to-use format, much like Fourier transforms or Ito calculus (see \cite{RMT} for an introductory account). One of the exciting, and relatively unexplored direction, is to add some financially motivated prior, like industrial sectors or groups, to improve upon the default ``agnostic” recipe.  

Now the data problem is solved as best as possible, the stationarity problem pops up. Correlations, like volatility, are not fixed in stone but evolve with time. Even the sign of correlations can suddenly flip, as was the case for the S\&P500/Treasuries during the 1997 Asian crisis. After 30 years of correlations staunchly in positive territory (1965 – 1997), bonds and equities have been in a ``flight-to-quality” mode  (i.e. equities down and bonds up) ever since. More subtle, but significant, changes of correlations can also be observed between single stocks and/or between sectors in the stock market. For example, a downward move of the S\&P500 leads to an increased average correlation between stocks. Here again, RMT provides powerful tools to describe the time evolution of the full covariance matrix \cite{RMT-dyn1, RMT-dyn2}. 

As I discussed in the previous section, stochastic volatility models have made significant progress recently, and now encode feedback loops that originate at the microstructural level. Unfortunately, we are very far from having a similar theoretical handle to understand correlation fluctuations, although Matthieu Wyart and I had proposed in 2007 a self-reflexive mechanism to account for correlation jumps like the one that took place in 1997 \cite{JEBO}. Parallel to the development of descriptive and predictive models, the introduction of standardized instruments that hedge against such correlation jumps would clearly serve a purpose. This is especially true in the current environment \cite{seager} where inflation fears could trigger another inversion of the equity/bond correlation structure, possibly devastating for many strategies that – implicitly or explicitly – rely on persistent negative correlations. Markowitz diversification free lunch can sometimes be poisonous!
 
\section{My Kingdom for a Copula}

As I just discussed, assessing linear correlations between financial assets is already hard enough. What about {\it non-linear} correlations then? If financial markets were kind enough to abide to Gaussian statistics, non-linear correlations would be entirely subsumed by linear ones. But this is not the case: genuine non-linear correlations pervade the financial world and are quite relevant, both for the buy side and the sell side. For example, tail correlations in equity markets (i.e. stocks plummeting simultaneously) are notoriously higher than bulk correlations. Another apposite context is the Gamma-risk of large option portfolios, the management of which requires an adequate description of quadratic return correlations of the underlying assets. 

In order to deal with non-linear correlations, mathematics has afforded us with a seemingly powerful tool -- ``copulas''. Copulas are supposed to encapsulate all possible forms of multivariate dependence.  But  in  the  zoo  of  all conceivable copulas, which one should one choose to faithfully represent financial data?

Following an unfortunate, but typical pattern of mathematical finance, the introduction of copulas twenty years ago has been followed by a calibration spree, with academics and financial engineers alike frantically looking for copulas to best represent their pet multivariate problem. But instead of first developing an intuitive understanding of the economic or financial mechanisms that suggest some particular dependence between assets and construct adequate copulas accordingly, the methodology has been to brute-force calibrate copulas straight out from statistics handbooks. The ``best'' copula is then decided from some quality-of-fit criterion, irrespective of whether the copula makes any intuitive sense at all. 

This is reminiscent of local volatility models for option markets: although the model makes no intuitive sense and cannot describe the actual dynamics of the underlying asset, it is versatile enough to allow the calibration of almost any option smile. Unfortunately, a blind calibration of some unwarranted model (even when the fit is perfect) is a recipe for disaster. If the underlying reality is not captured by the model, it will most likely derail in rough times — a particularly bad feature for risk management (recall the use of Gaussian copulas to price CDOs before the 2008 crisis). Another way to express this point is to use a Bayesian language: there are families of models for which the ‘prior’ likelihood is clearly extremely small, because no plausible scenarios for such models to emerge from market mechanisms. Statistical tests are not enough — the art of modelling is precisely to recognize that not all models are equally likely. 

The best way to foster intuition is to look at data before cobbling up a model, and come up with a few robust ``stylized facts'' that you deem relevant and that your model should capture. In the case of copulas, one interesting stylised fact is the way the probability that two assets have returns simultaneously smaller than their respective medians depends on  the linear correlation between the said two assets. Such a dependence exists clearly and persistently in stocks and, strikingly, it cannot be reproduced by most ``out-of-a-book'' copula families.
 
In particular, the popular class of so-called ``elliptical'' copulas is ruled out by such an observation. Elliptical copulas assume, in a nutshell, that there is a common volatility factor for all stocks: when the index becomes more or less volatile, all stocks follow suit. A moment of reflection reveals that this assumption is absurd, since one expects that volatility patterns are at least industry specific. But this consideration also suggests a way to build copulas specially adapted to financial markets. In Ref. \cite{Remy1}, R\'emy Chicheportiche and I showed how to weld the standard factor model for returns with a factor model for volatilities. Perhaps surprisingly, the common volatility factor is not the market volatility, although it contributes to it. With a relatively parcimonious parameterisation, most multivariate ``stylized facts'' of stock returns can be reproduced, including the non-trivial joint-probability pattern alluded to above.    

I have often ranted against the over-mathematisation of quant models, favouring theorems over intuition and convenient models over empirical data. Reliance on rigorous but misguided statistical tests is also plaguing the field. As an illustration related to the topic of copulas, let me consider the following question: is the univariate distribution of standardized stock returns {\it universal}, i.e. independent of the considered stock? In particular, is the famous ``inverse-cubic law'' \cite{cubic} for the tail of the distribution indeed common to all stocks? 

A standard procedure for rejecting such an hypothesis is the Kolmogorov-Smirnov (or Anderson-Darling) statistics. And lo and behold, the hypothesis is strongly rejected. But, wait -- the test is only valid if returns can be considered as independent, identically distributed random variables. Whereas returns are close to being uncorrelated, non-linear dependencies along the time axis are strong and long-ranged. Adapting the Kolmogorov-Smirnov test in the presence of long-ranged ``self-copulas'' is possible \cite{Remy2} and now leads to the conclusion that the universality hypothesis {\it cannot} be rejected. Here again, thinking about the problem before blindly applying standard recipes is of paramount importance to get it right.   

The finer we want to hone in on the subtleties of financial markets, the more we need to rely on making sense of empirical data, and to remember what the great Richard Feynman used to say: {\it It doesn't matter how beautiful your theory is, it doesn't matter how smart you are. If it doesn't agree with experiment, it's wrong.}

\section{Crowded Trades: Whales or Minnows?}

It is funny how, sometimes, seemingly obvious concepts do become paradoxical when one starts thinking harder about what they really mean. One topical example is the idea of ``crowded trades'' that has recently become a major talking point in the face of the disheartening performance of many Alternative Beta/Risk Premia funds. It seems self-evident to many that when investors pile into a given trade, future returns are necessarily degraded. But on the other hand, for each buyer there must be a seller -- so isn't the opposite trade crowded too? In what sense, then, is a crowded trade toxic? Can one come up with useful measures of crowding, that would allow one to construct portfolios as immune as possible to crowding risk?

In the mind of investors, the word ``crowding'' summons two distinct fears. One is that any mispricing that motivates a strategy is arbitraged away by the crowd, rendering that strategy  unprofitable in the future. The second is crash risk: harmful deleveraging spirals may occur as the crowd suddenly decides to run for the exit. Here we see how the symmetry between the two sides of the trade can be broken: a trade is crowded when investors on -- say -- the buy side are more prone to act in sync than those of the sell side. 

In fact, the most crowded trade of all is, and always has been, long the stock market. Crashes indeed happened many times in the past and will happen again in the future. Be that as it may, is investing in the stock market a bad decision? Certainly not, in fact all Risk Premia are profitable on the long run precisely because of such negatively skewed events. The equity risk premium is abnormally high -- but in fact it compensates for the deleveraging risk associated with the madness of crowds. In fact, we have argued in Ref. \cite{skewness} that the return of Risk Premia strategies are strongly correlated with their negative skewness, i.e. their propensity to crash. So in many cases, crowding is simply unavoidable -- the only question is whether the associated downside risk is adequately compensated or not. 

This brings us back to our first point: that of withering returns. The mechanism usually put forth is that the spread between the fundamental price and the market price is closed by those who trade based on that mispricing. If the cake is shared between a larger number of participants, each of them must get a smaller piece. Although this makes intuitive sense, the story cannot be that simple. When one looks at different measures of mispricing on which classical factors (momentum, profitability, low volatility, etc.) are supposed to thrive, there is no sign of a recent narrowing of the valuation spread between the short and long legs of the corresponding portfolios. The situation is even the exact opposite for price-to-book spreads, which have become wider since 2016 -- tantamount to saying that Value strategies are currently in the doldrums. 

But to argue that the plight of Value is due to crowding is at best misleading. Value as a strategy has been used extensively by market participants for decades. What we are seeing is more like a Value crash in slow motion, with investors getting out more aggressively of this strategy in the years 2019-2020, after a period of disappointing (but certainly not unprecedented!) performance. A rough estimation shows that if 250 B\$ are progressively redeemed from value strategies over a year, typical market neutral value portfolios should suffer a $\approx 20\%$ loss from price impact alone -- all very much in line with recent industry figures.

While there is no smoking gun that this is what happened recently, the argument that crowding is detrimental to convergent strategies (i.e. trades that reduce valuation spreads) is not without merit. But then how do we understand the effect of crowding on divergent strategies, such as momentum or trend following? Here, price impact arguments would suggest that more trend followers should bolster trends, not make them weaker. Crowding could even be beneficial for such strategies, at least up to a point.  

The problem with this optimistic surmise is that it neglects yet another facet of price impact, namely transaction costs. The point is that, according to our definition, crowded strategies are precisely those leading to correlated trades, i.e. many managers entering or leaving the market simultaneously. As an extreme outcome, this can lead to crashes, as discussed above. But even in perfectly normal regimes, trading in the same direction as others can significantly increase impact costs, a.k.a. ``slippage''. It is not my own traded quantity that matters, it is the aggregate quantity traded by all managers following the same trading idea. Although the strength of the trading signal is not necessarily impaired, crowded trades may suffer from so much ``co-impact'' \cite{co-impact} that the profitability of the strategy quickly shrivels to zero, or even below zero.        

This suggests an interesting metric to detect crowded strategies and estimate such co-impact  costs. The first step is to reconstruct the putative trades that a manager following a given strategy -- say Fama-French Momentum -- would send to the market on a given day. One then uses order-book tick data to determine the actual buy/sell order imbalances for each trading day. This allows one to compute the correlation of the overall market imbalance with the imbalance expected from the strategy under scrutiny. A statistically significant correlation means that the strategy does leave a detectable trace in the markets. One can also measure the correlation between these reconstructed trades and the price return of each stock; this provides a direct estimate of the co-impact costs. 

This is precisely what we did in Ref. \cite{crowded}. The conclusion is that the classical Fama-French momentum has indeed become more and more crowded in the last ten years, and, as of today, the estimated co-impact costs make the strategy all but unprofitable. The good news, on the other hand, is that there are many different ways to implement a given trading idea -- some more, some less correlated with the ``crowd''. This paves the way to portfolio constructions that attempt to minimize the correlation of trades with identified trading strategies, with the hope of eschewing the curse of crowded trades -- deleveraging spirals and all.     

\section{High-Frequency Trading \& Market Stability}

In the midst of the first COVID lockdown, the 10th anniversary of the infamous May 6th, 2010 ``Flash Crash'' went unnoticed. At the time, fingers were pointed at High Frequency Trading (HFT), accused of both rigging the markets and destabilizing them. Research has since then confirmed that HFT in fact results in significantly lower bid-ask spread costs and, after correcting for technological glitches and bugs, does {\it not} increase the frequency of large price jumps. In fact, recent models explain why market liquidity is intrinsically unstable: managing the risk associated to market-making, whether by humans or by computers, unavoidably creates destabilising feedback loops. In order to make markets more resilient, research should focus on better market design and/or smart regulation that nip nascent instabilities in the bud.  

Since orders to buy or to sell arrive at random times, financial markets are necessarily most of the time unbalanced. In such conditions, market-makers play a crucial role in allowing smooth trading and continuous prices. They act as liquidity buffers, that absorb any temporary surplus of buy orders or sell orders. Their reward for providing such a service is the bid-ask spread -- systematically buying a wee lower and selling a wee higher, and pocketing the difference. 

What is the fair value of the bid-ask spread? Well, it must at least compensate the cost of providing liquidity, which is {\it adverse selection}. Indeed, market-makers must post prices that can be picked up if deemed advantageous by traders with superior information. The classic Glosten-Milgrom model provides an elegant conceptual framework to rationalize the trade-off between adverse selection and bid-ask spread, but fails to give a quantitative, operational answer (see e.g. \cite{TQP} for a recent discussion). In a 2008 study \cite{spread} we came up with a remarkably simple answer: the fair value of the bid-ask spread is equal to the ratio of the volatility to the square-root of the trade frequency. This simple rule of thumb has many interesting consequences. 

First, it tells us that for a fixed level of volatility increasing the trade frequency allows market-makers to reduce the spread, and hence the trading costs for final investors. The logic is that trading smaller chunks more often reduces the risk of adverse selection. This explains in part the rise of HFT as modern market making, and the corresponding reduction of the spreads. Throughout the period 1900-1980, the spread on US stocks hovered around a whopping 60 basis points, whereas it is now only a few basis points. In the meantime, volatility has always wandered around $40 \%$ per year -- with of course troughs and occasional spikes, as we discuss below. In other words, investors were paying a much higher price for liquidity before HFT, in spite of wild claims that nowadays electronic markets are ``rigged''. In fact, after a few prosperous years before 2010, high frequency market-making has become extremely competitive and average spreads are now compressed to minimum values. 

From this point of view, the economic rents available to liquidity providers have greatly decreased since the advent of HFT. But has this made markets more stable, or has the decrease in the profitability of market-making also made them more fragile? The second consequence of our simple relation between spread and volatility relates to this important question. The point is that this relation can be understood in a two-way fashion: clearly, when volatility increases, the impact of adverse selection can be dire for market makers who mechanically increase their spreads. Periods of high volatility can however be quite profitable for HFT since competition for liquidity providing is then less fierce.

But in fact higher spreads by themselves lead to higher volatility, since transactions generate a larger price jump -- or even a crash when liquidity is low and the order book is sparse. So we diagnose a fundamental destabilising feedback loop, intrinsic to any market-making activity:
\[
\text{volatility} \quad  \longrightarrow  \quad \text{higher spreads \& lower liquidity}  \quad  \longrightarrow  \quad \text{more volatility}.
\]
Such a feedback loop can actually be included in stochastic order book models (such as the now commonly used family of ``Hawkes processes'' \cite{Hawkes_Review}). As the strength of the feedback increases, one finds a phase transition between a stable market and a market prone to {\it spontaneous liquidity crises}, even in the absence of exogenous shocks or news \cite{Fosset}.

This theoretical result suggests that when market-makers (humans or machines) react too strongly to unexpected events, liquidity can enter a death spiral. But who will blame them? As an old saying goes, {\it liquidity is a coward, it is never there when it is needed}.    

Such a paradigm allows one to understand why a large fraction of price jumps in fact occur without any significant news -- rather, they result from endogenous, unstable feedback loops \cite{News}. Empirically, the frequency of $10$-sigma daily moves of US stock prices has been fairly constant in the last 30 years, with no significant change between the pre-HFT epoch and more recent years \cite{TQP}. Even the 6th May 2010 Flash Crash has a pre-HFT counterpart: on the May 28th 1962, the stock market plunged 9\% within a matter of minutes, for no particular cause, before recovering -- much the same weird price trajectory as in 2010. Our conjecture: markets are intrinsically unstable, and have always been so. As noted in section 1 above, this chronic instability may lie at the heart of the turbulent, multiscale nature of financial fluctuations. 

Can one engineer a smart solution that make markets less prone to such dislocations? From our arguments above, we know that the task would be to crush the volatility/liquidity feedback loop, by promoting liquidity provision when it is on the verge of disappearing. One idea would be to introduce dynamical make/take fees, which would make cancellations more costly and limit order posting more profitable depending on the current state of the order book. These fees would then funnel into HFT's optimisation algorithms, and (hopefully) yank the system away from the regime of recurrent endogenous liquidity crisis.  

\section{Radical Complexity \& Scenario Based Macro-economics}

Good science is often associated with accurate, testable predictions. Classical economics has tried to conform to this standard, and developed an arsenal of methods to come up with precise numbers for next year's GDP, inflation and exchange rates, among (many) other things. Few, however, will disagree with the fact that the economy is a complex system, with a large number of heterogeneous interacting units, of different categories (firms, banks, households, public institutions) and very different sizes. In such complex systems, even qualitative predictions are hard. So maybe we should abandon our pretense of exactitude and turn to another way to do science, based on scenario identification. Aided by qualitative (agent based) simulations, swans that appear black to the myopic eye may in fact be perfectly white.   

The main issue in economics is precisely about the emergent organization, cooperation and coordination of a motley crowd of micro-units. Treating them as a unique representative firm or household clearly throws the baby with the bathwater. 
Understanding and characterizing such emergent properties is however difficult: genuine surprises can appear from micro- to macro-. One well-known example is the Schelling segregation model: even when all agents prefer to live is mixed neighborhoods, myopic dynamics can lead to completely segregated ghettos \cite{Schelling}. In this case, Adam Smith's invisible hand badly fails. 

More generally, slightly different micro-rules/micro parameters can lead to very different macro-states: this is the idea of ``phase transitions”; sudden discontinuities (aka crises) can appear when a parameter is only slightly changed.
Because of feedback loops of different signs, heterogeneities and non-linearities, these surprises are hard, if not impossible to imagine or anticipate, even aided with the best mathematical apparatus. 

This is what I would like to call ``Radical Complexity”. Simple models can lead to unknowable behaviour, where ``Black Swans'' or ``Unknown Unknowns'' can be present, even if all the rules of the model are known in detail. In these models, even probabilities are hard to pin down, and rationality is {\it de facto} limited. For example, the probability of rare events can be exponentially sensitive to the model parameters, and hence unknowable in practice \cite{PNAS}. In these circumstances, precise quantitative predictions are unreasonable. But this does not imply the demise of the scientific method. For such situations, one should  opt for a more qualitative, scenario based approach, with emphasis on mechanisms, feedback loops, etc. rather than on precise, but misleading numbers. This is actually the path taken by modern climate change science.

Establishing the list of possible (or plausible) scenarios is itself difficult. We need numerical simulations of Agent Based Models. While it is still cumbersome to experiment on large scale human systems (although more and more possible using web-based protocols), experimenting with Agent Based Models is easy and fun and indeed full of unexpected phenomena. These experiments {\it in silico} allow one to elicit scenarios that would be nearly impossible to imagine, because of said feedback loops and non-linearities. Think for example of the spontaneous synchronization of fireflies (or of neuron activity in our brains). It took nearly 70 years to come up with an explanation. Complex endogenous dynamics is pervasive, but hard to guess without appropriate tools. 

Experimenting with Agent Based Models is interesting on many counts. One hugely important aspect is, in my opinion, that it allows to teach students in a playful, engaging way how complex social and economic systems work. Such simulations would foster their intuition and their imagination, much like lab experiments train the intuition of physicists about the real world, beyond abstract mathematical formalism. 

Creating one’s own world and seeing how it unfolds clearly has tremendous pedagogical merits. It is also an intellectual exercise of genuine value: if we are not able to make sense of an emergent phenomenon within a world in which we set all the rules, how can we expect to be successful in the real world? We have to train our minds to grasp these collective phenomena and to understand how and why some scenarios can materialize and others not. The versatility of ABM allows one to include ingredients that are almost impossible to accommodate in classical economic models, and explore their impact on the dynamics of the systems \cite{Mark0,Mark0Covid}.  

ABM are often spurned because they are in general hard to calibrate, and therefore the numbers they spit out cannot and should not be taken at face value. They should rather be regarded as all-purpose {\it scenario generators}, allowing one to shape one's intuition about phenomena, to uncover different possibilities and reduce the realm of Black Swans. The latter are often the result of our lack of imagination or of the simplicity of our models, rather than being inherently impossible to foresee. 

Expanding the study of toy-models of economic complexity will create a useful corpus of scenario-based, qualitative macroeconomics \cite{Bookstater,Mounfield}. Instead of aiming for precise numerical predictions based on unrealistic assumptions, one should make sure that models rely on plausible causal mechanisms and encompass all plausible scenarios, even when these scenarios cannot be fully characterized mathematically. A qualitative approach to complexity economics should be high on the research agenda. As Keynes said, {\it it is better to be roughly right than exactly wrong.}   

\section*{Acknowledgments} I want to warmly thank all my collaborators on these topics, especially: R. Allez, R. Benichou, M. Benzaquen, P. Blanc, J. Bonart, F. Bucci, J. Bun, R. Chicheportiche, J. Donier, Z. Eisler, A. Fosset, M. Gould, S. Gualdi, S. Hardiman, A. Karami, Y. Lemp\'eri\`ere, F. Lillo, R. Marcaccioli, I. Mastromatteo, F. Morelli, M. Potters, P. A. Reigneron, P. Seager, D. Sharma, M. Tarzia, B. Toth, V. Volpati, M. Wyart, F. Zamponi.
I also want to pay tribute to various people with whom I had exciting and enlightening discussions on these matters, in particular R. Bookstaber, D. Farmer, X. Gabaix, J. Gatheral, J. Guyon, A. Kirman, C. Lehalle, J. Moran, M. Rosenbaum, N. Taleb. Finally, I am deeply indebted to Mauro Cesa who offered me the possibility of putting my thoughts together and publishing them as six monthly columns in Risk.net, from September 2020 to February 2021.

\end{document}